\documentstyle[twocolumn,pre,aps,epsbox]{revtex}

\begin{document}
\title{The Carnot Cycle for Small Systems: 
Irreversibility and the Cost of Operations}
\draft
\author{Ken Sekimoto}
\address{Yukawa Institute for Theoretical Physics, Kyoto University,
Kyoto 606-8502, Japan\\
E-mail: sekimoto@yukawa.kyoto-u.ac.jp
}
\author{Fumiko Takagi}
\address{Satellite Venture Business Laboratory,
        Ibaraki University,
        Ibaraki 316-8511, Japan\\
E-mail: fumiko@hcs.ibaraki.ac.jp
}
\author{Tsuyoshi Hondou}
\address{Department of Physics, Tohoku University,
Sendai 980-8578, Japan\\
E-mail: hondou@cmpt.phys.tohoku.ac.jp
}
\maketitle
\begin{abstract}
In the thermodynamic limit, the existence of a maximal
efficiency of energy conversion attainable by a Carnot cycle 
consisting of quasi-static isothermal and adiabatic processes 
precludes the existence of a perpetual machine of 
the second kind, whose cycles yield positive work 
in an isothermal environment.
We employ the recently developed framework of the energetics of
stochastic processes (called `stochastic energetics'),
to re-analyze the Carnot cycle in detail,
taking account of fluctuations, without taking 
the thermodynamic limit.
We find that both in this non-macroscopic situation, both 
processes of connection to and disconnection from
 heat baths and adiabatic processes that cause 
distortion of the energy distribution are sources of 
inevitable irreversibility within the cycle. 
Also, the so-called null-recurrence property 
of the cumulative efficiency of energy conversion over many cycles
and the irreversible property of isolated, purely mechanical
processes under external `macroscopic' operations are 
discussed in relation to the impossibility of a perpetual 
machine, or Maxwell's demon.
This analysis may serve as the basis for the design and analysis of
mesoscopic energy converters in the near future.
\end{abstract}
\pacs{05.90+m, 05.40-a,  05.70-a, 02.50-r}

\section{Introduction}
\label{sec:intro}
\subsection{Background}
The principles of thermodynamics were established in 
the last century as the universal laws characterizing the 
thermal and mechanical behavior of macroscopic systems.
The fact that we cannot control all the details of energy transfer
leads to the concept of {\it heat} as a form of energy flow, and the 
{\it Carnot cycle} has played a crucial role in the course
of investigation leading to the introduction of entropy as a state 
variable in addition to energy\cite{Fermi}.
On the other hand, Brownian motion and the stochastic dynamics 
of {meso}scopic systems in general have also been studied for many years,
and projection methods have allowed for the derivation of 
Langevin dynamics from microscopic Hamiltonian mechanics.
In a properly defined Langevin equation,  the influence of 
the unpredictable microscopic dynamics, which essentially 
represent the heat, is taken into account by Markovian random 
forces obeying the fluctuation-dissipation relationship.
In this manner, such an equation describes the canonical 
equilibrium distribution of the variables in question\cite{LandauSTD}.

Very recently, the concept of the heat on mesoscopic scales 
has been unambiguously defined in terms of Langevin dynamics\cite{ks1}.
We refer to the formalism providing this definition as
 {\it stochastic energetics}.%
The essential point of the thinking behind this formalism 
is that the heat transfered to the system is 
nothing but the microscopic work done by both frictional forces 
{\it and} the random force in the Langevin equation. 
The theoretical framework resulting from this realization 
widens the scope of application of Langevin dynamics to the extent that 
it can be used to describe not merely equilibrium states of
system in contact with heat baths, 
but also  general thermodynamic {\it processes} connecting 
different equilibrium states.
As a result, we can derive the first and the 
second laws of thermodynamics\cite{ks1,ks-ss} from 
stochastic energetics.
This formalism, together with projection methods, bridge 
a longstanding gap between microscopic Hamiltonian mechanics 
and macroscopic thermodynamics.
In this paper we apply the method of stochastic energetics to the investigation
of the Carnot cycle in the context of small systems.
To make this paper self-contained, we briefly summarize the 
framework of stochastic energetics in \S~\ref{sec:steng}.

Stochastic energetics has also been applied to the study of
thermodynamic processes under non-equilibrium conditions,
such as processes including two heat baths\cite{ks1} 
and processes in the presence of  steadily driving forces\cite{ks2}.
In particular, Feynman's ratchet model\cite{feynman} has been analyzed.
Regarding this model, the doubt has been cast by 
Parrondo\cite{parrondo}, and later by Sekimoto\cite{ks1} independently,
about the attainability of reversible energy
conversion with the `Carnot efficiency' $1-T_{\rm L}/T_{\rm H}$, where
$T_{\rm L}$ and $T_{\rm H}$  are the temperatures of cool and hot
heat baths. 
Analysis using stochastic energetics has shown 
explicitly that the efficiency of Feynman's ratchet is much less than
that of the Carnot efficiency mentioned above\cite{ks1}.

\subsection{Problems}
\label{subsec:problems}
With the descriptive power of stochastic energetics in hand,
we wish to reconsider the Carnot cycle.
We consider the Carnot cycle as an {\it object} of analysis 
within the theoretical framework of stochastic energetics.
Note that since we can derive the laws of thermodynamics 
directly using the stochastic energetics based on the Langevin
description, the role of the Carnot cycle in our study 
is {\it not} a source of theoretical results
from which one derives the laws of thermodynamics
as was its historical role.
(Of course both the Langevin description and 
thermodynamics have a microscopic basis in mechanics.)

We now describe our viewpoint in more detail.
Usually the Carnot heat engine is considered in an ideally macroscopic
context,working in the thermodynamic limit.
There, the small relative fluctuations of the variables, typically 
on the order of the inverse square root of the system size, are 
neglected.
Also, the cost involved in the operations of 
attaching/detaching the system under study 
 to/from heat baths is neglected, 
since this is not an the extensive quantity.
It is important to notice that the second law of thermodynamics, 
which is consistent with such a macroscopic Carnot engine, 
can exclude {\it only marginally} 
the existence of perpetual machine of the second kind
whose cycles yield positive work 
in an isothermal environment.
Thus we may gain a deeper insight into the nature of 
statistical thermodynamics and mechanics 
if we can formulate a method to take account of the finiteness of the 
the system under study 
as well as the cost involved in operations
of changing its interaction with heat baths,
in particular considering reversibility and the second law
of thermodynamics.

The approach of the present work is to construct the simplest 
model of the Carnot heat engine with a finite number (actually only three) 
of degrees of freedom, including the apparatus  connecting/disconnecting
it with heat baths, 
and to determine the effect of the finiteness of the system and the 
change resulting from operations of the type mentioned above.
As the system of study (or the `working material') we choose a single 
harmonic oscillator.
We show that there is an inevitable source of dissipation due to the 
{\it intrinsically} irreversible nature of the operations
of connecting and disconnecting it with heat baths,
and that, with the exception of such loss,
our model can attain the Carnot maximal efficiency defined 
as a properly defined average over infinitely many cycles, each of 
which is performed infinitely slowly.
At the same time this study reveals several basic problems which should
be further scrutinized in the future: 
one regards the smooth connection between 
the adiabatic process and the isothermal processes,
 and the other regards the irreversibility of adiabatic processes.
In the last section we  discuss these problems as well as the 
problem involving energy conversion with no help from external operations.

In the remaining part of this section 
we give a qualitative description of the aspects of the Carnot cycle
that we study in detail in the later sections.\\

\noindent
{\it 1.} {\sf The operations of connection to and disconnection from
 the heat bath.}\\
We ask first how we can describe {\it mechanically} the 
connection and disconnection of the system with the heat baths.
In an idealized picture, this basically consists of 
the switching on and off of the interaction between
the system and each heat bath.
In \S~\ref{sec:model} we describe explicitly a model that
realizes these operations.
We represent the influence of the heat baths by a 
frictional force and the random force of a Langevin equation, and 
we control the strength of the coupling between
the system and the heat baths by controlling the values of the corresponding 
interaction potentials.  We call these interaction
potentials `couplers'.
(In an actual mechanical system, 
the control of such a coupler could be exercised by a system of clutches.)

One could also imagine such control exercised
through the change of the friction constants
that appear in the Langevin equation.
In consideration of the absence of a definition of the required 
work to change these friction constants,
however, this idea is not pursued in the present paper.
.\\

\noindent
{\it 2.}{\sf  The reversible and irreversible work of operating
the couplers}\\
The operation of the couplers can, {\it in principle}, 
never be carried out quasi-statically, but, at the same time, that
 accompanying irreversible work can be made arbitrarily small.
The former part of this assertion is based on the following 
argument:
When the interaction between the system and a heat bath is 
strong, the energy transfer between them occurs with a 
short relaxation time. 
However, if we gradually weaken this interaction, this 
relaxation time increases more and more until it diverges when
the system is completely detached from the heat bath.
As long as the time-scale of the operation (i.e. the switching-off) is
finite, this operation can never remain  `slow' in comparison  
to the diverging relaxation time.
Thus the switching-off process is by no means quasi-static, or 
quasi-equilibrium.
(This is analogous to the non-adiabaticity encountered in
chemical reactions;
the Born-Oppenheimer approximation is inevitably invalid
when the distance between nuclei is neither sufficiently large
nor sufficiently small.)
Inevitable irreversibility can also exist in the process of strengthening
 the interaction between the system and the heat bath.
Such extreme strengthening of the interaction leads to the freezing of 
some degree(s) of freedom involved in the interaction, and
the mean first-passage time associated with these degrees of freedom
may become larger than the timescale of the operation (i.e. 
the strengthening).
In such a situation also, the operation can never be carried out 
quasi-statically.
Unlike the switching-off process, however, 
the indefinite strengthening of the interaction 
is not necessarily a part of the Carnot cycle. 
Despite this fact, we consider the latter
process in the sections that follow,
because this allows us to minimize the calculations
needed to reach a general conclusion.

We should, however, note that the inevitably irreversibile nature 
of the operations described above does not necessarily imply
an associated  large amount of irreversible work.
In \S~\ref{sec:clutch} we analyze the work involved in
 operating the coupler
and show  that the amount of irreversible work resulting from these
inevitably irreversible operations can be made arbitrarily small in 
the limit that the timescale of the operation becomes large. 
We also show that the reversible part of the work associated with these
operations remains finite in this limit, but that 
it cancels out within a cycle.\\

\noindent
{\it 3.}{\sf The condition for reversible contact between  
the system and a heat bath}\\
Temporarily putting aside the concept of irreversibility in the sense 
described in {\sf 2}
above, we can scrutinize the remaining part of the cycle and 
ask if and how the Carnot maximum efficiency can be 
attained.
With regard to a macroscopic Carnot cycle, 
according to textbook descriptions, 
in order to realize a reversible cycle,
`the temperature of the system should be the same as 
that of the heat bath with which 
the system is to make contact after an adiabatic process.' 
Strictly speaking, however, the energy, rather than the 
temperature, takes a definite value in a thermally isolated system, 
and the above statement needs to be  refined 
in terms of the language of probability. 
We argue in \S~\ref{sec:match} that reversible contact requires 
the probability distribution of the energy of the system 
just before interaction with a heat bath
to be identical to the canonical distribution 
at the temperature of the heat bath.
This condition can be satisfied if the system consists of harmonic 
oscillators, as the model described below. 
Generally, however, this is not the case, and 
in the general situation an irreversible process takes place 
when system contacts the heat bath, 
even though there occurs no net irreversible energy
transfer between the two (\S~\ref{subsec:1}).
In \S~\ref{sec:efficiencymax} we summarize the 
necessary conditions for Carnot cycle to realize 
the maximal efficiency $1-T_H/T_L$
without assuming the thermodynamic limit.
We show at the same time that this actually is the case for the model 
described in \S~\ref{sec:model}.\\

\noindent
{\it 4.} {\sf  Statistics of the efficiency of a finite number of cycles} \\
The efficiency of the energy conversion of an {\it individual} cycle is
 statistically distributed, because 
the energy possessed by the system is different each time 
the system is disconnected from a heat bath.
This fact reflects the indeterminate nature of the details of 
the microscopic states of the system and of the heat bath upon 
disconnection.%
As a consequence, if we define the cumulative `bonus' work as the 
difference between the cumulative work obtained over $n$ 
cycles and what we would expect from the Carnot maximal efficiency,
this bonus work takes the form of a discrete random walk as a function 
of $n$.
We show in \S~\ref{sec:efficiency} and the Appendix 
that the so-called null-recurrence
property of a one-dimensional random walk insures
that, although,
if we actually carry out a sequence of these cycles, the cumulative
bonus work we obtain will with probability 1 first become positive 
after a finite number of repetitions $n^*$, 
the statistical average of $n^*$ is infinite.

\noindent %
{\it 5.}  {\sf  Irreversibility of adiabatic processes} \\
The Carnot cycle includes an adiabatic process that is purely
mechanical.
We are interested in determining what work can be obtained through 
the cycle including {\it non}-quasi-static adiabatic processes.
If the efficiency in this case is increased 
in comparison to the quasi-static case,
the existence of a perpetual machine of the second kind is inspired, 
because our Carnot cycle can attain the 
maximal (reversible limit) efficiency under certain conditions 
specified in \S~\ref{sec:efficiencymax}.
In  \S~\ref{subsec:adiab} we show that, in relation to
the impossibility of a perpetual machine, there emerges the concept of 
the irreversibility of purely mechanical processes 
(with no assumption of the thermodynamic limit or mixing properties necessary)
under the influence of `macroscopic' operations by an external agent.
Here, designation of an operation as `macroscopic' implies that
(i) we are ignorant of the initial phase point 
on a given energy surface, and (ii) we 
are interested only in the 
statistical average over such an initial ensemble at a given energy.

\section{Brief summary of stochastic energetics}
\label{sec:steng}
We consider a Langevin equation that represents a system in contact 
with a heat bath at temperature $T$,
\begin{eqnarray}
\frac{dx}{dt}&=& \frac{p}{m}\cr
\frac{dp}{dt}&=&-\gamma \frac{p}{m}-\frac{\partial U(x,a)}{\partial x}+\xi(t).
\label{eq:Langevin2}
\end{eqnarray}
Here we denote by $x$ and $p$ the dynamical variable of the system and
its conjugate momentum, while $m$ represents the mass,
$\gamma$ the friction constant, and $U$ the potential energy for $x$.
We assume that $U$ may depend on, in addition to $x$, 
the variable (or variables) $a$, 
which is controlled by an external agent (or agents).
The function $\xi(t)$ represents, as usual, Gaussian white noise 
obeying the relations (hereafter we adopt units in which $k_{\rm B}=1$)
\begin{equation} \langle\xi(t)\rangle=0,
\qquad\langle\xi(t)\xi(t^\prime)\rangle=2\gamma T\delta(t-t^\prime).
\end{equation}
The second relation (Einstein's relation) insures a
canonical distribution of $x$ and $p$ at temperature $T$
if the parameter $a$ is held fixed for an infinitely long time.

Multiplication of each term in the second equation in 
(\ref{eq:Langevin2}) by the displacement $dx$ 
yields the equation\cite{gardiner} 
\begin{equation}
\frac{d}{dt}\left(\frac{p^2}{2m}\right)dt
=\left(-\gamma\frac{p}{m}+\xi(t)\right)dx
-\frac{\partial U(x,a)}{\partial x}dx,
\label{eq:balance2}
\end{equation}
where we have used the first equation of Eq.(\ref{eq:Langevin2}) and 
also the identity $\frac{dp}{dt}dx=\frac{dp}{dt}\frac{p}{m}dt$.
We note that $-\left(-\gamma \frac{dx}{dt}+ \xi(t)\right)$  
is the {\it reaction} force  exerted by the system against the heat bath,
since the frictional force $-\gamma\frac{dx}{dt}$ and 
the random force $\xi(t)$ are both due to %
the heat bath. 
We identify the work done by the reaction force
as the  {\it heat} transferred from the system to the heat bath, 
which we denote by $(-d{  Q})$\cite{ks1}:
\begin{equation}
-d{ Q} \equiv -\left(-\gamma\frac{dx}{dt}+\xi(t)\right)dx.
\end{equation}
(The minus sign in front of ${ d Q}$ is included to conform to 
the convention of thermodynamics textbooks.)
The key point of introducing the concept of heat is that, 
although the heat bath is idealized and not affected by 
the system's dynamics, 
the heat bath can still be subject to a reaction force exerted by
the system.
Adding the total differential $dU$ to both sides 
of Eq.(\ref{eq:balance2}),
we obtain the general expression for the {\it energy balance} as
\begin{equation}
d\left(\frac{p^2}{2m}+U\right) =\frac{\partial U}{\partial a}da+d{  Q}.
\label{eq:first}
\end{equation}
Now, because the l.h.s. is the total increase of the energy, 
and $d{  Q}$ is the energy input to the system as a heat,
the first term on the r.h.s. of Eq.(\ref{eq:first}) must be 
identified as the {\it work} done by the external system, $dW$, on 
the system through the change of the variable $a$, 
\begin{equation}
dW \equiv \frac{\partial U}{\partial a}da.
\label{eq:work}
\end{equation}
We conclude that the law of energy balance expressed as
\begin{equation}
dE=dW+d{  Q}, \qquad E\equiv \frac{p^2}{2m}+U
\label{eq:balance}
\end{equation}
is satisfied for any {\it single} realization of the 
stochastic process described by Eq.(\ref{eq:Langevin2}).

For a quasi-static process, in which $|\frac{da}{dt}|$ is arbitrarily 
small, the work is reversible and is equal to the change in the Helmholtz 
free energy $F(T,a)$ with probability 1. 
That is, in an ensemble of
infinitely many realizations of such a process, the probability distribution
of the work becomes a point distribution concentrated at the value
of $F(T,a)$.
\begin{equation}
dW =dF(T,a), \qquad (\mbox{for a quasi-static process with $T$ fixed}),
\label{eq:second0}
\end{equation}
with
\begin{equation}
F(T,a)\equiv  -T \log \left[ \int\!\!\!\int 
e^{-\frac{E}{T}}{\rm d}x{\rm d}p\right]
.
\label{eq:helmholtz}
\end{equation}
The derivation of Eq.(\ref{eq:second0}) is as follows.
We first note that, for $a$ to change
by any small but finite amount $da$, it takes a time 
$|da|/|\frac{da}{dt}|$, 
which is indefinitely large in the quasi-static limit.
During this time interval
the state point $(x,p)$ comes arbitrarily close to almost all 
possible values, and its empirical distribution 
becomes asymptotically equal to
the canonical distribution, 
$P_{\rm eq}(x;T,a)=\exp\left[\frac{F(T,a)-E}{T}\right].$
(The exception here is the case in which the 
interval $[a,a+da]$ includes a point at which 
the equilibration time diverges. See {\sf 2} of
\S~\ref{subsec:problems} and \S~\ref{sec:clutch} below.)
We can then evaluate $\frac{\partial U}{\partial a}da$ using its
average with respect to $P_{\rm eq}$ in the quasi-static limit.
Using the identity,
\begin{equation}
\int dx  \frac{\partial U(x,a)}{\partial a}P_{\rm eq}(x;T,a)=
\frac{\partial F(T,a)}{\partial a} \quad\mbox{($T$ fixed)},
\end{equation}
we reach the result Eq.(\ref{eq:second0}).

In fact Eq.(\ref{eq:second0}) is a stronger statement than the usual
second law of thermodynamics for extensive systems,
Note that for a thermodynamic system constituted by an 
ensemble of a large number of independent stochastic systems 
obeying Eq.(\ref{eq:Langevin2}),
the first law of thermodynamics is obtained from Eq.(\ref{eq:balance}),
\begin{equation}
\left\langle dE\right\rangle
 =\langle dW\rangle +\langle d{  Q}\rangle, 
\label{eq:firstav}
\end{equation}
and the second law of thermodynamics for quasi-static 
processes is obtained from Eq.(\ref{eq:second0})\cite{ks-ss,chris},
\begin{equation}
\langle dW\rangle =dF(T,a), \qquad 
(\mbox{for a quasi-static process with $T$ fixed}).
\label{eq:second}
\end{equation}
These relations are concerned with only 
the ensemble averages denoted by $\langle \cdot \rangle$.
It has also been shown\cite{ks-ss} that, for a finite rate of 
change of $a(t)$, the Clausius inequality
\begin{equation}
\langle dW\rangle \ge dF(T,a)
\label{eq:second2}
\end{equation}
holds, and an explicit formula for the {\it irreversible} work,
$\langle dW\rangle -dF(T,a)$, has been obtained up to the second order 
in $da/dt$.

\section{Model}
\label{sec:model}
\begin{figure}[b]
\postscriptbox{7.80cm}{3.66cm}{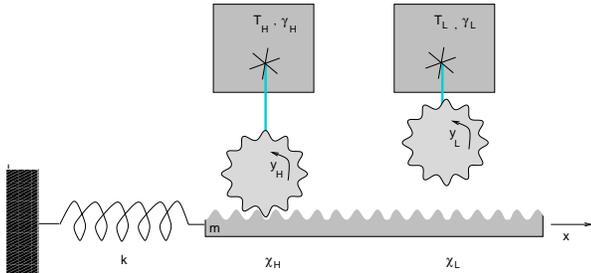}
\caption{Schematic view of a Carnot heat engine.
The spring and the shaded linear `gear' represent the harmonic oscillator 
as the ``system.'' 
The left end of the spring (the black box) is fixed.
Heat baths of temperatures $T_{\rm H}$ and $T_{\rm L}$ (the 
square shaded boxes) exert forces 
on the vanes (the star-shaped symbols inside the heat baths) 
whose angles of rotation are denoted by
$y_{\rm H}$ and $y_{\rm L}$, respectively.
These vanes are tightly connected to the circular gears.
These circular gears can interact with the system
in a manner that depends on the control parameters $\chi_{\rm H}$ 
and $\chi_{\rm L}$ of the couplers.}
\label{fig:notion}
\end{figure}

Figure \ref{fig:notion} schematizes the idea of our model.
We employ a single harmonic oscillator with mass 
$m$ and spring constant $k\, (>0)$ as the system under study, 
which we call simply the ``system''.
We denote by $x$ and $p$ the position and momentum of the system.
correspond to compressing [decompressing] the ideal gas.
Below, we consider
 $k$ to be a quantity that can be controlled, as the volume of a gas
system is controlled in macroscopic Carnot cycles.

In order to allow independent and variable interaction with each heat 
bath, we represent  each such interaction in the form of a mechanical 
force, which subsumes the corresponding frictional and  
Gaussian random forces.
Such mechanical forces should be related in some way to the
degrees of freedom that directly interact with the heat baths,
which we denote by $y_{\rm H}$ and $y_{\rm L}$.
For simplicity, we do this by writing the mechanical forces
as interaction forces,
$-\frac{\partial \phi_{\rm H}}{\partial x}$ and 
$-\frac{\partial \phi_{\rm L}}{\partial x}$.
As interaction potentials, we choose functions 
$\phi_{\rm H}(x-y_{\rm H},\chi_{\rm H})$ and 
$\phi_{\rm L}(x-y_{\rm L},\chi_{\rm L})$, where
$\chi_{\rm H}$ and $\chi_{\rm L}$ are the control
parameters. We call $\phi_{\rm H}$ and $\phi_{\rm L}$ the `couplers',
because their values directly indicate the strength of the coupling between
the system and the respective heat baths. We use the expressions like
`control the coupler(s)' in reference to changes made in the values of
these control parameters.
We assume that the functions
$\phi_\alpha$ ($\alpha$=H, L) 
are  $2\pi$-periodic functions  of $x-y_\alpha$. (See, for details,
\S~\ref{sec:clutch} and Fig.~\ref{fig:clutch}.)
\noindent
The degrees of freedom $y_{\rm H}$ and $y_{\rm L}$ are subject to
the frictional forces $-\gamma_{\rm H} \frac{dy_{\rm H}}{dt}$
 and $-\gamma_{\rm L} \frac{dy_{\rm L}}{dt}$ and the random
forces $\xi_{\rm H}(t)$ and $\xi_{\rm L}(t)$ exerted by the 
heat baths at temperatures $T_{\rm H}$ and $T_{\rm L}$,
respectively, 
as well as the interaction forces from the system, 
$-\partial \phi_{\rm H}/\partial y_{\rm H}$
and $-\partial \phi_{\rm L}/\partial y_{\rm L}$.
Here, $\gamma_{\rm H}$ and $\gamma_{\rm L}$ are the 
friction constants, and $\xi_{\rm H}(t)$ and $\xi_{\rm L}(t)$ 
are the white Gaussian random forces satisfying 
$\langle \xi_{\rm{H}}(t)\xi_{\rm{H}}(t^\prime)\rangle =
 2 \gamma_{\rm{H}} T_{\rm{H}} \delta(t-t^\prime)$, $\langle \xi_{\rm{L}}(t)\xi_
{\rm{L}}(t^\prime)\rangle = 2 \gamma_{\rm{L}} T_{\rm{L}} \delta(t-t^\prime)$, a
nd $\langle \xi_{\rm{H}}(t)\xi_{\rm{L}}(t^\prime)\rangle = 0$.
The equations of motion for $x$, $p$, $y_{\rm H}$, and $y_{\rm L}$ are
given as follows:
 \begin{mathletters}
 \label{eq_of_mot}%
 \begin{equation}
 \frac{dx}{dt}=\frac{p}{m},
 \label{equationx}
 \end{equation}
 \begin{equation}
 \frac{dp}{dt}= -kx-\frac{\partial \phi_{\rm H}}{\partial x}
 -\frac{\partial \phi_{\rm L}}{\partial x},
 \label{equationp}
 \end{equation}
 \begin{equation}
 \gamma_{\rm H}\frac{dy_{\rm H}}{dt}
 =-\frac{\partial \phi_{\rm H}}{\partial y_{\rm H}}+\xi_{\rm H}(t),
 \label{equationyH}
 \end{equation}
 \begin{equation}
 \gamma_{\rm L}\frac{dy_{\rm L}}{dt}
 =-\frac{\partial \phi_{\rm L}}{\partial y_{\rm L}}+\xi_{\rm L}(t).
 \label{equationyL}
 \end{equation}
 \end{mathletters}
We consider the gears of the heat bath and the system to be `tightly
connected' (i.e. completely engaged) for $\chi_\alpha =1$, that is,
the interaction $\phi_\alpha(x-y_{\alpha},1)$ is so
strong that the difference $x-y_{\alpha}$ is fixed except for a small
thermal fluctuation around its mean value, 
while these gears are `disconnected' (i.e. completely disengaged)
for $\chi_{\alpha}=0$, that is,
$\phi_\alpha(x-y_{\alpha},0)\equiv 0.$
We have neglected the inertia effect related to $y_{\rm H}$ 
and $y_{\rm L}$, as they would play only secondary role for 
our analysis.

\begin{figure}[ht]
 \hspace*{2cm}
    \postscriptbox{5.23cm}{6.66cm}{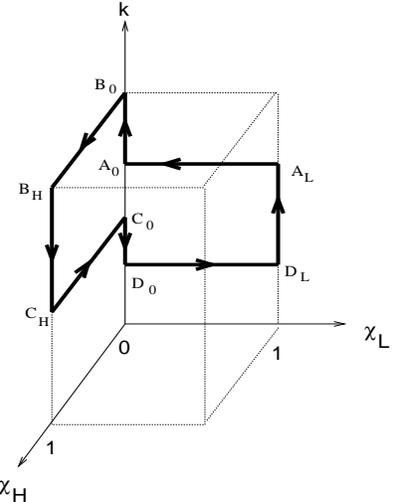}
 \caption{The cycle undergone by the control parameters.
The various legs of this cycle correspond
to the following processes of the system:
 isothermal processes
(${\rm B}_{\rm H}\to {\rm C}_{\rm H}$ and 
${\rm D}_{\rm L}\to {\rm A}_{\rm L}$),  
 adiabatic processes 
(${\rm A}_{\rm 0}\to {\rm B}_{\rm 0}$ and 
${\rm C}_{\rm 0}\to {\rm D}_{\rm 0}$),  
and the remaining processes where
only one of the two control parameters of the couplers 
is changed (i.e. $\chi_{\rm H}\neq 0$ 
exclusively or  $\chi_{\rm L}\neq 0$)
while $k$ is kept constant.
}
 \label{fig:protocol}
 \end{figure}

The protocol by which we control the parameters 
is represented in the space of $(k, \chi_{\rm H}, \chi_{\rm L})$
as shown in Fig. \ref{fig:protocol}.
\noindent In the figure, the paths along the axis 
$(\chi_{\rm H},\chi_{\rm L})=(0,0)$ 
correspond to the adiabatic processes, while the vertical paths with 
$(\chi_{\rm H},\chi_{\rm L})=(1,0)$ and 
$(\chi_{\rm H},\chi_{\rm L})=(0,1)$ 
correspond to the isothermal processes.
The values of $k$ corresponding to these four horizontal paths
are the parameters.

\section{Reversible and irreversible work of operating the couplers}
\label{sec:clutch}
As we discussed in \ref{sec:intro}, the operation of the couplers can 
never be made quasi-static, because the time-scale of this
operation inevitably becomes shorter than the equilibration
time for the system when the coupling between
the system and a heat bath becomes either absent or 
extremely tight.
In addition to the work due to these irreversible processes, there is
also reversible work associated with operation of coupler.

\begin{figure}[ht]
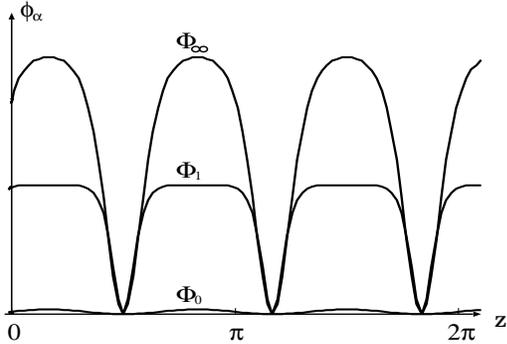

\hspace*{0.5cm}
   \postscriptbox{6.73cm}{4.66cm}{clutch.eps}
\caption{The profiles of the interaction potential $\phi_\alpha$ are 
given as functions of $z\equiv x-y_\alpha$ for the three typical
values of the maximum of $\phi_\alpha$, 
 $\Phi_0$, $\Phi_1$ and $\Phi_\infty$, 
where $\Phi_0 \ll T_\alpha \ll \Phi_1 < \Phi_\infty$
(see the text).
}
\label{fig:clutch}
\end{figure}

Figure~\ref{fig:clutch} illustrates the generic features 
of the potential $\phi_\alpha(z,\chi_\alpha)$ 
for three different values of $\chi_\alpha$.
Here $\Phi_0$, $\Phi_1$, and $\Phi_\infty$ represent the
height of each potential profile.
We assume that the height of $\phi_\alpha(z,\chi_\alpha)$ is 
a monotonically increasing function of $\chi_\alpha$ and
satisfies $\max_z \phi(z,0)=0$, 
$\max_z \phi(z,\chi_{\alpha 0})=\Phi_0 \,(\ll T_\alpha)$, 
$\max_z \phi(z,\chi_{\alpha 1})=\Phi_1 \,(\gg T_\alpha)$, and
$\max_z \phi(z,1)=\Phi_\infty \,(>\Phi_1)$ with
$0 < \chi_{\alpha 0}< \chi_{\alpha 1}< 1$.

There are two situations in which
the time-scale of measurement and/or operation 
cannot exceed the equilibration time of the system. 
One is when the height of $\phi_\alpha$ is very small, and
the other is when the height of $\phi_\alpha$ is very large.
Let us assume that the regime $0<\chi_\alpha <\chi_{\alpha 0}$ 
corresponds to the former case; that is,
for $\max_z \phi_\alpha (z,\chi_\alpha) \le \Phi_0$, 
the interaction $\phi_\alpha$ is so weak that 
the equilibration time of the system with the heat bath ($T=T_\alpha$)
is beyond the timescale of measurement and/or operation. 
We call this  
the {\it loose regime}.
Then, we assume that the regime $\chi_{\alpha 1}<\chi_\alpha <1$ 
corresponds to the latter case; that is,
for $\max_z \phi_\alpha (z,\chi_\alpha) \ge \Phi_1$, 
the interaction $\phi_\alpha$ is so strong that 
the equilibration time 
characterized by the over-barrier transition of $z$ 
(see Fig~\ref{fig:clutch}) is again beyond the 
timescale of measurement and/or operation. 
In particular, we assume that for $\chi_\alpha =1$ there occur
essentially no thermal activation events over 
the barrier $\Phi_\infty$.
We call this 
the {\it tight regime}.
In the remaining regime,
$\chi_{\alpha 0}\leq \chi_\alpha \leq \chi_{\alpha 1}$ we assume
that the operation can be carried out in a manner that
arbitrarily closely approximates the quasi-static limit.
Note that what Kramers calls the 
`small viscosity' and `large viscosity' cases in Ref.\cite{kramers}
correspond, respectively, to 
the limits of the loose and tight regimes.

We now evaluate the work %
\begin{equation}
 W_\alpha(\chi_{\alpha 1}\!\to \!\chi_{\alpha 2})\equiv  \!\!\!
\int_{\chi_\alpha=\chi_{\alpha 1}}^{\chi_{\alpha 2}}
\!\!\!\!\!\!
\frac{\partial \phi_{\alpha}(x(t)\!- \! y_\alpha(t),\chi_\alpha(t))
}{\partial \chi_{\alpha}}  d\chi_{\alpha}(t),
\label{eq:clutchwork}
\end{equation}
for the loose and tight regimes using Eq.(\ref{eq:work}).
We will describe the case of $\alpha=$H for concreteness.
(The case of $\alpha=$L can be treated similarly.)
In the loose regime (the processes near 
B$_0$ and C$_0$ in Fig.~\ref{fig:protocol}), 
$W_{\rm H}(0\to \chi_{{\rm H}0})$ is 
of order $\Phi_0$ and is small $(\ll T_{\rm H})$,
although this work may be mostly irreversible.
Since $\Phi_0$
is associated with the lower limit of quasi-static operation, 
the timescale of the operation is required to be large enough
to satisfy the condition $\Phi_0 \ll T_{\rm H}$. 

In the tight regime (the processes near 
$B_{\rm H}$ and $C_{\rm H}$ in Fig.~\ref{fig:protocol}), 
the situation is more subtle.
The contribution to 
$W_{\rm H}(\chi_{{\rm H}1}\to 1)$ 
consists of {\it (i)} the contribution produced when $z$ moves
around the valley regions of $\phi_{\rm H}(z,\chi_{\rm H})$ with
$\phi_{\rm H}\lesssim \Phi_1$ and
{\it (ii)} the contribution 
produced when $z$ visits, by 
 {\it rare} thermal excitation, the barrier regions of
$\phi_{\rm H}$. 
To simplify the analysis we exclude the former 
contribution by assuming that
$\partial \phi_{\rm H}(z,\chi_{\rm H})/\partial \chi_{\rm H}=0$ for 
$z$ in the valley regions of $\phi_{\rm H}$ (Fig.~\ref{fig:clutch}) in
the tight regime.
(This assumption is only technical; one can reach the conclusion
of this paragraph without it.)
The evaluation of the contribution {\it (ii)} above is carried out 
as follows.
The probability to find $z$ in the barrier region is  
$\sim e^{-{\Phi_1}/{T}}$, 
and for such values of $z$, 
the change of $\chi_{\rm H}$ from $\chi_{{\rm H}1}$ to 1 
results in an amount of work $\sim (\Phi_\infty -\Phi_1)$. 
Thus we have $W_{\rm H}(\chi_{{\rm H}1}\to 1)$ 
$\sim e^{-{\Phi_1}/{T}}(\Phi_\infty -\Phi_1)
\sim e^{-{\Phi_1}/{T}}\Phi_\infty $.
Because the timescale of operation is sufficiently large to
allow large values of $\Phi_1$.
For this reason, the conditions 
$e^{-{\Phi_1}/{T}}\Phi_\infty \ll T$ and $\Phi_\infty \gg T$ 
can be satisfied simultaneously.
In conclusion, the irreversible part of the 
work associated with both the loose 
regime and the tight regime can be made as small as we wish by 
making the timescale of the operation sufficiently long in these regimes.
The same conclusion holds for the case of $\alpha={\rm L}$.

The quasi-static work associated with 
the change of $\chi_{\alpha}$ within 
the region $\chi_{\alpha 0}\le \chi_{\alpha}\le \chi_{\alpha 1}$
can be evaluated using Eq.(\ref{eq:second0}).
Below we show that such quasi-static work cancels exactly 
when summed over the consecutive operations of 
connection and disconnection with a heat bath.
Again, considering the case of $\alpha={\rm H}$,  we denote by
$F(T_{\rm H},k,\chi_{\rm H},0(=\chi_{\rm L}))$ 
the Helmholtz free energy of the 
composite system of the harmonic oscillator and 
the couplers, 
$\{p, x, y_{\rm H},y_{\rm L}\}$:
\begin{eqnarray}
 & &e^{- F(T_{\rm H},k,\chi_{\rm H},0)/{T_{\rm H}}}
 \equiv 2\pi \int_{-\infty}^\infty dp 
 \int_{-\infty}^\infty dx \int_{0}^{2\pi} dy_{\rm H}\cr
 & &
 \exp\left\{ -\frac{1}{T_{\rm H}} \left [ 
 \frac{p^2}{2m}+\frac{kx^2}{2}+\phi_{\rm H}(x-y_{\rm H},\chi_{\rm H})
 \right ]\right\}.
 \label{helmholtz}
 \end{eqnarray}
Note that here $\chi_{\rm L}=0$ and 
the factor $2\pi$ in front of the integration on the r.h.s. comes
from the phase integration over $y_{\rm L}$.
Performing the integration over $y_{\rm H}$ first, we have
\begin{equation}
F(T_{\rm H},k,\chi_{\rm H},0)=
-\frac{T_{\rm H}}{2}\log
\frac{ {(2\pi)}^4 {(T_{\rm H})}^2 m}{k}
+\tilde{F}(T_{\rm H},\chi_{\rm H},0),
\label{eq:clutchF}
\end{equation}
with $\tilde{F}$ defined by 
\begin{equation}
e^{-\tilde{F}(T_{\rm H},\chi_{\rm H},0)/T_{\rm H}}=
\int_{0}^{2\pi} dz
 e^{-{\phi_{\rm H}(z,\chi_{\rm H})}/{T_{\rm H}}}.
\label{eq:etildeF}
\end{equation}
The first term on the r.h.s of Eq.(\ref{eq:clutchF})
is independent of $\chi_{\rm H}$, while 
$\tilde{F}$ is independent of  $k$. 
Using the notation of Eq.(\ref{eq:clutchwork}),
we find from Eq.(\ref{eq:clutchF}) that 
$W_{\rm H}(\chi_{{\rm H}0}\to \chi_{{\rm H}1})=$
$\tilde{F}(T_{\rm H},\chi_{{\rm H}1},0)-
\tilde{F}(T_{\rm H},\chi_{{\rm H}0},0)$
along B$_0 \to$B$_{\rm H}$ in Fig.~\ref{fig:protocol}, and
$W_{\rm H}(\chi_{{\rm H}1}\to\chi_{{\rm H}0})=$
$\tilde{F}(T_{\rm H},\chi_{{\rm H}0},0)-
\tilde{F}(T_{\rm H},\chi_{{\rm H}1},0)$ 
along C$_{\rm H} \to$C$_0$.  
These two cancel exactly:
\begin{equation}
W_{\rm H}(\chi_{{\rm H}0}\to \chi_{{\rm H}1})+
W_{\rm H}(\chi_{{\rm H}1}\to \chi_{{\rm H}0}) =0.
\label{eq:workadBC}
\end{equation}
The actual time interval required 
to change $\chi_\alpha$ between $\chi_{\alpha 0}$
and $\chi_{\alpha 1}$ is finite,  say,  $t_{\rm 01}$.
The irreversible work due to this finiteness has been shown to be 
${\cal O}({t_{\rm 01}}^{-1})$ quite generally\cite{ks-ss}.
Thus the irreversible work associated with the process in which
$\chi_\alpha$ is changed between $\chi_{\alpha 0}$ and
$\chi_{\alpha 1}$ can be made as small as we wish by 
making the timescale of operation sufficiently long.

For later use, we now also estimate the heat exchanged 
upon the operation of the couplers.
As we have shown above, the amount of work 
in the loose and tight regimes can be made arbitrarily small. 
Also, changes in the parameters $\chi_\alpha$ lead 
to only small changes of the internal energy of the composite system.
These facts together with the energy balance principle (see 
Eq.(\ref{eq:balance})) lead to the conclusion that the heat exchanged 
in these two regimes can be made as small as we wish.
Next, the heat exchanged during the quasi-equilibrium processes with 
 $\chi_{\alpha 0}<\chi<\chi_{\alpha 1}$
is assessed as follows.
From Eq.(\ref{eq:clutchF}) the ensemble average of the internal energy of 
the composite system is given by
$T_{\rm H}+(1-T_{\rm H}\frac{\partial}{\partial T_{\rm H}})
\tilde{F}(T_{\rm H},\chi_{\rm H},0)$.
If we define by $\langle Q({\rm B}_0\to {\rm B}_{\rm H})\rangle$
the average heat influx to the composite system during
the quasi-equilibrium operation along B$_0\to$B$_{\rm H},$ 
the condition of energy balance, Eq.(\ref{eq:firstav}), yields
 $\langle Q({\rm B}_0\to {\rm B}_{\rm H})\rangle=
-T_{\rm H}\frac{\partial}{\partial T_{\rm H}}
[\tilde{F}(T_{\rm H},\chi_{{\rm H}1},0)
-\tilde{F}(T_{\rm H},\chi_{{\rm H}0},0)].$
This heat cancels exactly the average heat input 
$\langle Q({\rm C}_{\rm H}\to {\rm C}_0)\rangle$
similarly defined along C$_{\rm H}\to$C$_0$:
\begin{equation}
\langle Q({\rm B}_0\to {\rm B}_{\rm H})\rangle 
+\langle Q({\rm C}_{\rm H}\to {\rm C}_0)\rangle =0.
\label{eq:heatadBC}
\end{equation}
In the same manner,
we can show the cancellation of both the work and the heat 
during the quasi-static part of the operation of coupler
along D$_0 \to$D$_{\rm L}$ and A$_{\rm L} \to$A$_0$.

\section{Matching the `temperature' of the system and a heat bath}
\label{sec:match}

Here we study the meaning of the idea of matching the `temperature' 
of the small system with that of a heat bath.
For comparison, we note that this meaning is unambiguous
for an isolated macroscopic thermodynamic system, for which the 
energy and the temperature are simultaneously well-defined
quantities, and  in order to realize a reversible Carnot cycle,
the temperature of the system should be the same as 
that of the heat bath with which it makes contact.
Contrastingly, for isolated small systems,
the energy takes a definite value, while the temperature
is not generally well-defined.
We will show in this section that if the small system is a
harmonic oscillator, the concept of the temperature is still useful,
and reversibility can be obtained as in macroscopic systems.
Discussion regarding the general case is given in 
\S~\ref{subsec:1}.

Suppose that a coupler is operated quasi-statically up to
the edge of a loose regime ($\chi_{\rm L}=\chi_{{\rm L}0}$
 along ${\rm A}_{\rm L}\to {\rm A}_0$ or
$\chi_{\rm H}=\chi_{{\rm H}0}$ along
${\rm C}_{\rm H}\to {\rm C}_0$ in Fig.~\ref{fig:protocol}).
The energy of the oscillator in this situation  
fluctuates as a function of time and though the temporal fluctuation of 
the energy is very slow, it still obeys the canonical distribution 
at the temperature of the heat bath ($T_{\rm L}$ for 
A$_0$ and $T_{\rm H}$ for C$_0$) 
up to a small error of ${\cal O}(\Phi_0)\, (\ll T_\alpha)$.
By the definition of the loose regime, further weakening the connection
results in a situation in which there is no 
appreciable exchange of energy 
between the system and the heat bath
(see \S~\ref{sec:intro} and \ref{sec:clutch}).
Then, the complete disconnection leaves an
isolated system whose energy is distributed according 
to the canonical distribution corresponding 
to the temperature of the heat bath.
This is in fact true even if the small system is not
a harmonic oscillator.

For a harmonic oscillator, however,
both the energy distribution in the canonical ensemble and the
transformation of the system's energy through 
the quasi-static adiabatic process have special features.
The energy distribution in the canonical ensemble 
at temperature $T$, 
$P_{\rm can}(E;T)$ is independent of the parameter $k$:
\begin{equation}
P_{\rm can}(E;T)=\frac{1}{T}e^{-\frac{E}{T}}.
\label{eq:canonical}
\end{equation}
(See Appendix~\ref{sec:derivation} for a derivation.)
Then, at A$_0$ and C$_0$, the energy of the oscillator
is distributed according to $P_{\rm can}(E;T_{\rm L})$ 
and $P_{\rm can}(E;T_{\rm H})$, respectively.
If the oscillator has a specific (initial) energy $E$ and
undergoes the quasi-static adiabatic process represented as
A$_0\to$B$_{\rm 0}$ or C$_0\to$D$_{\rm 0}$ in Fig.~\ref{fig:protocol},
its energy, $E(k)$, changes so that the 
value $J(E,k)$ given by (\ref{eq:harmonicJ}) remains constant 
\cite{landau:cm}:
\begin{equation}  
J(E(k),k)= \frac{E(k)}{2\pi}\sqrt{\frac{m}{k}}=\mbox{constant}.
\label{invariant}
\end{equation}
The relation (\ref{invariant}) determines the change of the 
energy distribution
when the system undergoes a quasi-static adiabatic process through the 
change of $k$:
For a change from $k$ to $k'$, the altered distribution $P'(E')$
must obey $P_{\rm can}(E,T)dE=P'(E')dE'$ with 
$E'/({2\pi\sqrt{\frac{k'}{m}}})=E/({2\pi\sqrt{\frac{k}{m}}})$.
Thus we obtain the energy distribution of a canonical ensemble 
{\it{after}} the change in $k$,
$P^\prime (E')=P_{\rm can}(E',T')$
with $T'$ being defined by 
$T/\sqrt{k}=T'/\sqrt{k'}$.
However, we note that this simple situation is due to the special nature
of the harmonic oscillator system.
The general case is discussed in \S~\ref{subsec:1}.

With these facts in mind, we can now characterize the condition for a
quasi-equilibrium transition between adiabatic processes and
subsequent isothermal processes:
\begin{description}
\item[(1)] The energy distribution of the system before the connection
with the heat bath must be that of a canonical ensemble.
\item[(2)] The temperature characterizing this canonical ensemble
must be the same as that of the heat bath in question.
\end{description}
The reason is that, under these conditions, the system upon 
connection behaves statistically 
{\it as if} it had been in contact 
with the heat bath for a long time.
Note that the connection should be begun 
through a sufficiently weak interaction 
$\sim \Phi_0$ (see \S~\ref{sec:clutch}).
In our case with the protocol described by 
Fig.~\ref{fig:protocol}, we require
that $T'=T_{\rm H}$ at B$_0$ and $T'=T_{\rm L}$ at D$_0$
hold, respectively.
Denoting the value of $k$ at 
that between $A_0$ and $A_{\rm{L}}$ as $k_A$,
that between $B_0$ and $B_{\rm{H}}$ as $k_B$, etc., the condition
for a quasi-equilibrium connection to the heat bath
is given explicitly as follows:
\begin{equation}
\frac{T_{\rm L}}{\sqrt{k_{\rm A}}}=\frac{T_{\rm H}}{\sqrt{k_{\rm B}}}, 
\quad
\frac{T_{\rm H}}{\sqrt{k_{\rm C}}}=\frac{T_{\rm L}}{\sqrt{k_{\rm D}}}.
\label{matching1}
\end{equation}
\section{Efficiency}
\label{sec:efficiencygen}
\subsection{How is the Carnot limit approached?}
\label{sec:efficiencymax}              

We now evaluate the maximal
overall efficiency. We must take into account
(i) the operation of the couplers, (ii) the isothermal processes,
and (iii) the adiabatic processes.

(i) We assume that by making the time in the 
loose and tight regimes sufficiently long,
the intrinsically irreversible  work and  heat flow 
can be made as small as we wish. 
The remaining part of the operation of couplers 
is assumed to be made
under quasi-equilibrium conditions.  The accompanying
work and heat flow cancel exactly when summed over
an infinite number of consecutive connection 
and disconnection with the heat bath
 (see Eqs.(\ref{eq:workadBC}) and (\ref{eq:heatadBC})).

(ii) For the isothermal parts of the cycle,
B$_{\rm H}\to$C$_{\rm H}$ and D$_{\rm L}\to$A$_{\rm L}$,
we assume a quasi-static change of $k$.
The accompanying work is then 
given using the general formula Eq.(\ref{eq:second0}).
For the part B$_{\rm H}\to$C$_{\rm H}$, 
$F(T,a)$ in Eq.(\ref{eq:second0}) is replaced by 
$F(T_{\rm H},k,\chi_{\rm H},0)$ 
of Eq.(\ref{eq:clutchF}), and the work done by 
the system is
$ \frac{T_{\rm H}}{2} \log\left(\frac{k_{\rm B}}{k_{\rm C}}\right),$ 
which we denote by $-W({\rm B}_{\rm H}\to {\rm C}_{\rm H})$.
Similarly, for the part D$_{\rm L}\to$A$_{\rm L}$ the work is
$-W({\rm D}_{\rm L}\to {\rm A}_{\rm L})
  =  \frac{T_{\rm L}}{2} \log\left(\frac{k_{\rm D}}{k_{\rm A}}\right).$
Then, the relation (\ref{matching1}), we have
\begin{equation}
-W({\rm B}_{\rm H}\to {\rm C}_{\rm H})-
W({\rm D}_{\rm L}\to {\rm A}_{\rm L}))=\frac{T_{\rm H}-T_{\rm L}}{2} 
\log\left(\frac{k_{\rm B}}{k_{\rm C}}\right).
\label{sumBCDA}
\end{equation}

(iii) 
For the adiabatic part of the cycle, 
A$_{\rm 0}\to$B$_{\rm 0}$ and C$_{\rm 0}\to$D$_{\rm 0}$,
we also assume a quasi-static change of $k$.
The energy of the system then 
obeys the law (\ref{invariant}) and the amounts of work
done by the system in the adiabatic processes
$-W({\rm A}_{\rm 0}\to {\rm B}_{\rm 0})$ and 
$-W({\rm C}_{\rm 0}\to {\rm D}_{\rm 0})$ 
are given by
$-W({\rm A}_{\rm 0}\to {\rm B}_{\rm 0})=%
E(k_{\rm A})(1-\sqrt{k_{\rm B}/k_{\rm A}})$ and
$-W({\rm C}_{\rm 0}\to {\rm D}_{\rm 0})=%
E(k_{\rm C})(1-\sqrt{k_{\rm D}/k_{\rm C}}).$
As the energies $E_{\rm A}$ and $E_{\rm C}$
obey the distribution Eq.(\ref{eq:canonical}) with $T=T_{\rm L}$ and
$T=T_{\rm H}$, respectively, 
their statistical averages are 
$\langle E_{\rm A}\rangle =T_{\rm L}$ and 
$\langle E_{\rm C}\rangle =T_{\rm H}$ 
[up to a small error of ${\cal O}(\Phi_0)$]. 
Using (\ref{matching1}), we then have 
\begin{eqnarray}
&& -\langle W({\rm A}_{\rm 0}\to {\rm B}_{\rm 0})\rangle -
\langle W({\rm C}_{\rm 0}\to {\rm D}_{\rm 0})\rangle \cr
&&=   T_{\rm L}\left[1-\sqrt{\frac{k_{\rm B}}{k_{\rm A}}}\right]+
T_{\rm H}\left[1-\sqrt{\frac{k_{\rm D}}{k_{\rm C}}}\right]=0.
\label{sumCD}
\end{eqnarray}
While we obtain this simple result in the present case, it is important
to note that the cancellation of 
the contributions from the adiabatic processes on the 
average is not a generic feature of the Carnot cycle  
(consider, for example, non-ideal gases).

The heat influx from the high temperature heat bath is evaluated as
follows.
While the work during the isothermal quasi-equilibrium processes 
is a non-fluctuating quantity  
(see Eq.(\ref{eq:second0})), both the energy influx
from the heat bath and the system's energy fluctuate 
subject to the constraint of energy balance described by
Eq.(\ref{eq:balance}).
From Eqs.(\ref{eq:clutchF}) and (\ref{eq:etildeF}) we can show that 
the average internal energy of the composite system 
with degrees of freedom $\{p, x, y_{\rm H},y_{\rm L}\}$
is independent of the parameter $k$. 
Therefore, the statistical average of the heat influx during the 
isothermal process ${\rm B}_{\rm H}\to{\rm C}_{\rm H}$, which we 
denote by $\langle Q({\rm B}_{\rm H}\to{\rm C}_{\rm H})\rangle$, 
satisfies 
\begin{equation}
0= \langle Q({\rm B}_{\rm H}\to{\rm C}_{\rm H})\rangle
+W({\rm B}_{\rm H}\to {\rm C}_{\rm H}).
\label{eq:energy0}
\end{equation}
Thus we have
\begin{equation}
\langle Q({\rm B}_{\rm H}\to{\rm C}_{\rm H})\rangle
=\frac{T_{\rm H}}{2} \log\left(\frac{k_{\rm B}}{k_{\rm C}}\right).
\label{eq:input}
\end{equation}
As we have seen in \S~\ref{sec:clutch}, there is no net heat flow due 
to the quasi-static part of the operation of coupler 
(see Eq.(\ref{eq:heatadBC})), while the heat transfer associated with 
the loose and tight regimes can be made as small as we wish.

Collecting the above results, the maximal overall 
efficiency $\eta_{\rm max}$ of the cycles is reduced to 
the following formula:
\begin{equation}
\eta_{\rm max}=-\frac{W({\rm B}_{\rm H}\to {\rm C}_{\rm H})-
W({\rm D}_{\rm L}\to {\rm A}_{\rm L})
}{\langle Q({\rm B}_{\rm H}\to{\rm C}_{\rm H})\rangle}. 
\label{eq:eta}
\end{equation}
Then, using (\ref{sumBCDA}) and (\ref{eq:input}) we have
\begin{equation}
\eta_{\rm max}= 1-\frac{T_{\rm L}}{T_{\rm H}}.
\label{eq:etamax}
\end{equation}

We would like to stress that the attainment of this efficiency
 (whose expression is familiar from textbook treatments) is not
 due to the quasi-static operation of the whole system.
In the situation we consider, we 
have seen that some parts of the cycle can never be carried out
quasi-statically, due to the intrinsically irreversible 
operation of the couplers  (\S~\ref{sec:clutch}), as well as the intrinsic 
irreversibility resulting from the non-canonical energy 
distribution of the systems caused by the adiabatic 
processes(\S~\ref{sec:match}).

\subsection{Statistics over a finite number of cycles}
\label{sec:efficiency}
The maximal efficiency Eq.(\ref{eq:etamax}) obtained above
represents exclusively the ratio of the total work done by the 
system to the total energy influx from the 
high temperature heat bath through infinite number of
cycles.
Here we consider the efficiency for a single cycle,
$\eta_{\rm loc}$, which can be written as
\begin{equation}
\eta_{\rm loc}=\frac{\frac{T_{\rm H}-T_{\rm L}}{2} 
\log\left(\frac{k_{\rm B}}{k_{\rm C}}\right)-\delta_W}{
\frac{T_{\rm H}}{2} 
\log\left(\frac{k_{\rm B}}{k_{\rm C}}\right)+\delta_Q},
\label{eq:etaloc}
\end{equation}
with
\[-\delta_W=
- W({\rm A}_{\rm 0}\to {\rm B}_{\rm 0}) -
 W({\rm C}_{\rm 0}\to {\rm D}_{\rm 0}),\]
and
\begin{eqnarray}
\delta_Q &=& 
\left\{Q({\rm B}_0\to {\rm B}_{\rm H})
+ Q({\rm C}_{\rm H}\to {\rm C}_0)\right\}
\cr
&&+\left\{
Q({\rm B}_{\rm H}\to{\rm C}_{\rm H})-
\frac{T_{\rm H}}{2} 
\log\left(\frac{k_{\rm B}}{k_{\rm C}}\right)
\right\}.
\end{eqnarray}
We first note several properties of $\eta_{\rm loc}.$

{\it 1.}~The deviations $\delta_W$ and $\delta_Q$
do not vanish, even in the quasi-equilibrium limit, since
the system continues to exchange energy with a heat bath
until the moment that it is disconnected 
from the heat bath.

{\it 2.} If we choose the initial point of an individual cycle to be
somewhere between D$_{\rm L}$ and A$_{\rm L}$, 
then the values of $\eta_{\rm loc}-\langle \eta_{\rm loc}\rangle$ 
for different cycles are statistically independent.
In fact, the statistical deviations of 
$W({\rm A}_{\rm 0}\to {\rm B}_{\rm 0})$ and 
$W({\rm C}_{\rm 0}\to {\rm D}_{\rm 0})$ are mutually
uncorrelated because of the intervening isothermal Markov
processes ${\rm B}_{\rm H}\to{\rm C}_{\rm H}$ and 
${\rm D}_{\rm L}\to{\rm A}_{\rm L}$,
while $W({\rm A}_{\rm 0}\to {\rm B}_{\rm 0})$ and 
$Q({\rm B}_{\rm 0}\to{\rm B}_{\rm H})$ are statistically
correlated through the shared point B$_0$, 
as are $Q({\rm C}_{\rm H}\to{\rm C}_{\rm 0})$ and 
$W({\rm C}_{\rm 0}\to {\rm D}_{\rm 0})$ through 
the point C$_0$. 

{\it 3.} One can define the `excess output', 
$-\stackrel{\circ}{W}$, by
\begin{eqnarray}
-\stackrel{\circ}{W}&\equiv&
(\eta_{\rm loc}-\eta_{\rm max})\left[{\frac{T_{\rm H}}{2} 
\log\left(\frac{k_{\rm B}}{k_{\rm C}}\right)+\delta_Q}\right] \cr
&=& -\delta_W-\eta_{\rm max}\,\delta_Q.
\label{eq:bonus}
\end{eqnarray}
A positive value of $-\stackrel{\circ}{W}$ implies 
that we happened to get more work than that expected from 
the Carnot maximal efficiency (i.e., $-\delta_W>\eta_{\rm max}\,\delta_Q$).
Such a situation can result through fluctuations,
and it is not in contradiction
with the second law of thermodynamics.
We may then, however, ask how many cycles on average
we must carry out before we first obtain a cumulative excess
output,
$-\stackrel{\circ}{w}_n\equiv \sum_{i=1}^n (-\stackrel{\circ}{W}_i)$, where 
$-\stackrel{\circ}{W}_i$ is the excess output of the 
$i$-th cycle and $n$ is the total number of consecutive cycles.
The point of this question can be understood in terms of the 
following apparent paradox:
Suppose one monitors $-\stackrel{\circ}{w}_n$ as a function of $n$
and stops when it becomes positive for the first time.
If one could repeat such a procedure of monitor-and-stop indefinitely
many times, one could construct a perpetual machine of the second kind.
This, of course, would be in contradiction with the second law 
of thermodynamics.
A pitfall of this false argument is that,
although for a given sequence of cycles, the condition
$-\stackrel{\circ}{w}_n>0$ will be satisfied at some finite $n$ with
probability 1, the {\it average} over separate sequences
of the smallest value of $n$ with positive  
$-\stackrel{\circ}{w}_n$ is divergent.
This fact is closely related to the fact that the 
one-dimensional random walk is `null-recurrent'  (see 
Appendix~\ref{sec:nullrecurrent}).


\section{Discussion}
\label{sec:conclusion}
\subsection{Irreversibility resulting from contact with a heat bath}
\label{subsec:1}
In \S~\ref{sec:match} we have used the fact that
for the harmonic oscillator system, as a result of the
quasi-static adiabatic process, the energy changes 
in such a manner that the energy distribution
remains in canonical form, with simply a 
change of the temperature, $T\to T'$.
However, the harmonic oscillator represents a special system,
and this is not generically the case with any Hamilton.
More generally, the energy distribution $P_{\rm can}(E,T)$ is distorted 
into some non-canonical form, $P'(E)$, as a result of 
the quasi-static adiabatic process.
When an ensemble of systems following the distribution $P'(E)$
are brought into (weak) contact 
with a heat bath of arbitrary temperature $T'$, 
the energy distribution {\it irreversibly} relaxes to the canonical
form $P_{\rm can}(E,T')$. 
This is the case even if $T'$ is 
chosen so that $\int EP'(E)dE =\int E P_{\rm can}(E,T')dE$, 
i.e. even in the case that 
no net heat is transferred on the average from the bath to the system.
The relevance of this irreversible relaxation 
to the energetics of small systems requires further scrutiny.
This will be discussed in more detail in a separate paper\cite{SSHT}.

\subsection{Irreversible adiabatic process}
\label{subsec:adiab}
The adiabatic process in the Carnot cycle is a mechanical process.
The ergodic invariant 
theorem\cite{kasuga}
 tells us that under
quasi-static and adiabatic change of a system parameter, say $a$,
by a finite amount,
the phase volume enclosed by the energy surface defined by 
the system's energy at each moment, $J(E,a)$ (see (\ref{eq:defJ})
for definition), remains constant.
The theorem does not assume the thermodynamic limit nor
the presence or absence of chaotic trajectories.
Contrastingly, for a non-quasi-static process
$J(E,a)$ can either increase or decrease,
depending on both the nature of the change of $a$  and the 
initial conditions of the system.

In the context of the present paper,
 however, it is most meaningful to confine ourselves to only
`macroscopic' external operations, excluding  `demonic' ones which
depend on detailed information of the system. 
More precisely, we focus on an unprejudiced choice of 
the initial conditions among those with a given energy, and also
focus only on  statistical averages of 
the energy, rather than consider
particular results obtained from particular initial conditions.
Sato\cite{ksato} has recently studied a harmonic oscillator under a
time-dependent force,   %
  $ m\ddot{x}=-x +a(t)$, 
as the simplest non-trivial example of a system with only 
macroscopic external operations.
Here, the change of $a(t)$ amounts to a horizontal displacement 
of the potential. He showed analytically, as one can easily confirm, 
that for an arbitrary function $a(t)$, 
the energy of the oscillator, 
$\frac{m}{2}\dot{x}^2+\frac{1}{2}(x-a)^2$, 
is strictly non-decreasing {\it if} 
the average is taken 
with respect to the initial condition over all initial conditions
represented by states with a given energy, i.e., over
a micro-canonical ensemble.
Only in the limit of a quasi-static process $(|\frac{da}{dt}|\to 0)$
is the energy unchanged.

This example demonstrates the irreversibility of a 
mechanical system of non-macroscopic size 
 {with}\, properly defined macroscopic operation.
A recent numerical work\cite{komatsu2} investigating a harmonic 
oscillator with time-dependent spring constant, 
$ m\ddot{x}=-k(t)x$, 
reveals the same phenomena when $k(t)$ is constrained to return
to its initial value.
In the present context of the analysis of the efficiency of 
Carnot cycle (\S~\ref{sec:efficiencygen}),
these findings are both natural and important, since, if the 
case were different,
one could find functional form for $k(t)$ 
for which the adiabatic work on the system would be
less than what we expect for a quasi-static process, and the 
whole cycle could be used to construct a perpetual machine of the
second kind.
(Note that in the analysis of \S~\ref{sec:efficiencygen} 
we have excluded {only marginally} the existence of such a
perpetual machine if the loss due to the other 
sources is made arbitrarily small.)

It is desirable to obtain
a general proof (or counter-evidence) 
of the irreversibility resulting from non-quasi-static processes.
A related analysis using path probabilities has been performed for  
chaotic systems\cite{sasa00}.
However, we suspect that 
the essential mechanism of the irreversibility related to the 
characterization of macroscopic operators, which are ignorant of 
the initial microscopic state of the system, can be elucidated
without resort to chaotic statistics.
%
It is also a future problem to scrutinize the case in which 
the topology of the energy contour surface 
in the phase space (${\cal H}_a=E$) changes 
at some parameter value, say $a_c$. 
In such cases, quasi-static processes cannot be extended 
across $a_c$\cite{8figure}.

\subsection{Control of processes by the system itself}
\label{subsec:3}
As in the case of the ordinary macroscopic Carnot cycle, 
we have introduced control parameters, 
$\{k,\chi_{\rm H},\chi_{\rm L}\}$.
We assumed that the values of these parameters are 
changed by some external agent whose dynamics are external to
the equation of motion of the system.

In fact, however, there are many `self-controlled' energy transducers
which contain their own control systems. 
In such cases, the identification of the control system is more or less
a matter of interpretation.
In the macroscopic world,
DC electric motors and steam engines are examples, 
while motor proteins, such as myosins, kinesins 
and dyneins, etc.,  are microscopic examples.
Theoretically, the so-called Feynman ratchet and pawl system\cite{feynman}
has been proposed as a microscopic 
energy transducer working by itself between hot and cool heat baths.
In this model, the role of the control system is played either
by the pawl or by the ratchet, depending on which of these two 
is in direct contact with the cool heat bath. The stochastic 
energetics of this model have been analyzed\cite{ks1,magnasco98}.
B\"utticker's model\cite{buttiker} is another self-controlled
microscopic transducer.
In this model a massive particle moves while in contact with a heat
bath of position-dependent temperature, $T(x)=T_{\rm H}$  or  $T_{\rm L}$.
In this model, the inertia of the particle serves to switch the
particle's environment from $T=T_{\rm H}$ to $T=T_{\rm L}$, or vice versa.
The stochastic energetics of this model have also 
been analyzed\cite{mm-ss,hondou-ks}.
Some people have claimed that the Carnot limiting efficiency
$\eta_{\rm max}= 1-\frac{T_{\rm L}}{T_{\rm H}}$ can be attained in
Feynman's ratchet and pawl system
(see \cite{feynman} and \cite{sakaguchi}) and in 
B\"utticker's model (see \cite{astumian}).
With the exception of the original work by Feynman\cite{feynman}, where
no implementation details are given, 
these studies introduced into their analyses some 
`gate' mechanism.  
The study of the energetics of such systems, including 
the action of these gates, has not yet been made.

\acknowledgements
Discussions with K. Sato, S. Sasa, T. Komatsu and  T. Chawan-ya 
on the adiabatic processes of harmonic oscillator are gratefully
acknowledged.  The helpful comments by G. Paquette on the manuscript
are also acknowledged. This work is supported in part 
by the Inamori Foundation (T.H.), and a Grant in Aid by the 
Ministry of Education, Culture and Science (Priority Area,
Nos.11156216 and 12030215) (K.S.).

\appendix
\section{Derivation of (21) }
\label{sec:derivation}
For a general Hamiltonian ${\cal H}_a$ with a parameter $a$, 
the energy distribution $P_{\rm can}^{{\cal H}_a}(E;T)$ 
corresponding to the canonical ensemble at temperature T is 
\begin{equation}
P_{\rm can}^{{\cal H}_a}(E;T,a)=W(E,a)e^{\frac{F(T,a)-E}{T}},
\label{eq:PcanGeneral}
\end{equation}
with 
\[
W(E,a)\equiv\frac{\partial J(E,a)}{\partial E},
\] 
\begin{equation}
J(E,a)\equiv \int_{E>{{\cal H}_a}} d\Gamma, 
\label{eq:defJ}
\end{equation}
\[ 
e^{-\frac{F(T,a)}{T}}\equiv \int e^{-\frac{{\cal H}_a}{T}}d\Gamma,
\] 
where $\int d\Gamma$ denotes 
the phase integral. 
Here, $J$,  or $S\equiv \log J$, is an adiabatic invariant.

In the text, ${\cal H}_a$ is that of an isolated harmonic 
oscillator, $\frac{p^2}{2m}+\frac{k x^2}{2}$,
and we take its spring constant $k$ as $a$.
The calculation of $J(E,k)$ is straightforward, yielding 
\begin{equation}
J(E,k)=  \frac{E}{2\pi}\sqrt{\frac{m}{k}}.
\label{eq:harmonicJ}
\end{equation}
$W(E,a)$ is, therefore, independent of $E$. Thus from 
(\ref{eq:PcanGeneral}) we reach (\ref{eq:canonical}).
In general, however, $W(E,a)$ depends on $E$, and 
$P_{\rm can}^{{\cal H}_a}(E;T)$ 
is not simply an exponential $\sim e^{-E/T}$.

\section{Null-reccurence property}
\label{sec:nullrecurrent}
As $\{-\stackrel{\circ}{W}_i\}$ are statistically independent
of each other,  $-\stackrel{\circ}{w}_n$ constitutes
a one-dimensional discrete random walk.
To simplify the argument we assume that $-\stackrel{\circ}{W}_i$
takes only the values $\pm 1$ randomly.
If we denote by $f_{2n}$ the probability that 
at the $(2n-1)$-th step that the random walker comes to 
the position $+1$ for the first time,  
it is known that 
\[f_{2n}=            
\frac{1}{n 2^{2n+1}} \left(
\begin{array}{c}
2n \\ 
n 
\end{array}
\right)
\simeq \frac{1}{\sqrt{4\pi}n^{{3}/{2}}}.\] 
The fact that $f_{2n}$ is normalized ($\sum_{n=1}^\infty f_{2n}=1$)
implies that this event occurs 
with probability 1 at some $n$. 
On the other hand, it is also true that
\[\sum_{n=1}^\infty (2n-1) f_{2n}=\infty.\] 
which is referred to as the null-recurrence property.
%
Thus if we are to wait until the position of the walker becomes positive
for the $M$-th time, with $M\ge 1$, then
the `waiting time' is, {\it on the average}, infinite.

%

%
%


\begin{references}
\bibitem{chris}  
 We have noticed that the relation 
(\protect\ref{eq:second0}) to hold without averaging over the ensemble
thanks to discussion with Chris Jarzynski.
Until now, only its weak form (\protect\ref{eq:second}) 
had been shown (see Ref.\protect\cite{ks-ss}).
\end{references}
\end{document}